\documentclass[preprint,psfig]{aastex}

\shorttitle{Afterglow of GRB000301c}
\shortauthors{Bhargavi \& Cowsik}
\begin{document}

\title{Early Observations of the Afterglow of GRB000301c}


\author{S. G. Bhargavi\altaffilmark{1} and R. Cowsik\altaffilmark{1,2}}
\altaffiltext{1}{Indian Institute of Astrophysics, Bangalore, 560 034, India.}
\altaffiltext{2}{Tata Institute of Fundamental Research, Mumbai, 400 005, India.}
\email{bhargavi@iiap.ernet.in ;  cowsik@iiap.ernet.in}

\begin{abstract}
We report multiband observations of the Optical Transient (OT)
associated with GRB000301c carried out between 2--4 March 2000
using the 2.34-m Vainu Bappu Telescope (VBT) at Kavalur, India.
When combined with other reported data, the initial decline in the R-band
magnitude with log ($t-t_0$), the time since the burst is fit with a slope
$\alpha_{1}$ = -0.70 $\pm$ 0.07 which steepens after about 6.0 days to a
slope of $\alpha_{2}$ = -2.44 $\pm$ 0.29. This change in slope does not 
occur smoothly but there is an indication for a bimodal distribution.
The available measurements of the evolution of (B--R) color do not show any
discernible evolution in the first 12 days.
\end{abstract}
 

\section{Introduction}

Ever since in a pioneering effort, Costa (1997) and Van Paradijs (1997)
detected the first ever counterpart, associated with GRB970228,
identification of the optical counterparts of Gamma Ray Bursts (GRB) and
rapid follow-up observations in other wavelength bands have given impetus to
GRB astronomy during the last 3 years. 
The fading counterpart of a GRB, also known as 'Afterglow', is generally monitored
down to the faintest detection limit of a telescope
to study its brightness variation as a function of time since the occurence
of the burst event,
to derive the decay law from the light curve and to try to locate precisely
the burst counterpart with respect to the host galaxy, if any has been identified.
A study of all the afterglows so far detected shows a diversity of light 
curve properties
and it is necessary to accumulate a statistically significant set of such
observations for the classification of the bursts and for subsequent
theoretical modelling. Two excellent reviews by Kulkarni et al. (2000) and by
Piran (1999) provide a clear picture of the current status of this rapidly
evolving field of astronomy.
 
GRB000301c was detected by All Sky Monitor (ASM) on board {\it Rossi X-Ray 
Timing Explorer (RXTE) }
on 2000, March 1.4108UT and also by two other spacecrafts {\it Ulysses} and
{\it NEAR}.
The burst had a single peak with slow decay structure lasting 10 seconds and
was localized by an error box of area 50 sq. arcmin centred
at (J2000.) RA=$16^{h}20^{m}21^{s}.5$, DEC=+29$^{\circ}24^{\prime}56^{\prime
\prime}.37$ (Smith, Hurley \& Cline 2000).
The Optical counterpart was first reported by Fynbo et al.(2000a) and the 
redshift has
been measured to be z=1.95 $\pm$ 0.1 by Smette et al.(2000) using HST and to
be z=2.03 $\pm$ 0.003 by Castro et al. (2000) using the Keck-II telescope.
The OT was visible at R=27.9$\pm$0.15 from the late-time HST imaging on
19 Apr 2000 (Fruchter et al. 2000a), but there was no evidence for a host galaxy
underlying the GRB to a magnitude $\sim$ 28.5.
 
In the next section we present the details of our observations, data 
analysis and results. In section 3 we discuss briefly our conclusions.

\section{Observations, Analysis and Results}

At the 2.34m Vainu Bappu Telescope (VBT), the follow-up observations in the 
B, R and I bands \footnote{CCD images are available at 
{\it http://www.iiap.ernet.in/new{\_}results/GRB000301c.html}}
of GRB000301c began 3 hours after the e-mail notification of the burst event
on March 2, 2000. We could continue the observations on March 3 and 4 also
as the telescope time was already allotted for our ongoing program on GRBs.
Our observations
were carried out with a Tek 1024 CCD with $24\mu\times24\mu$ pixels positioned
at the f/3.25 prime focus of the telescope covering a FOV of 
about $10^{\arcmin}\times
10^{\arcmin}$ with an image scale of 0.$^{\arcsec}$61/pixel.
The sky conditions were clear with an average seeing of 2.$^{\arcsec}$5; however,
the quality of the acquired data was found to be better on March 2 and 4 compared
to those obtained on March 3. 
The details of the observations along with the magnitudes are 
summarized in Table 1.

The pre-processing of CCD frames
viz. de-biasing, flat-fielding and removal of cosmic rays is accomplished in
a standard
manner using the IRAF software package. Thereafter the images taken with the same
filters are co-added after aligning them.
The magnitudes were determined at an aperture of
$7^{\prime\prime}.2$ with aperture corrections applied for fainter objects including
the OT. Since the average seeing was $2.^{\prime\prime}5$ in some of
our R-band exposures the OT is blended with the nearby star 'A' of Garnavich(2000)
at a distance of about $6^{\prime\prime}$.
In such instances we have also estimated the magnitude of OT by masking the
nearby bright star after fitting a circularly symmetrical Gaussian profile and 
obtained consistent results.
 
In the photometric reduction standard methods have been applied and the magnitudes
thus estimated have uncertainties less than 0.01m and the values agree with those
provided by Henden(2000) for some select stars.

\section{Discussions \& Conclusions}

In the preceeding sections we presented our observations and analysis of GRB000301c
afterglow observations using Vainu Bappu Telescope, Kavalur, India.
In order to place our observations in the context
of other observations in the R-band we have collected the data available through GCN
circulars and those given by Jensen et al. (2000) in Table.~2. who reported
the most extensive coverage
of the event. Following the standard practice in combining the photometric data
obtained by different groups and different instruments, we renormalize all 
measurements to that of Jensen et al. (2000). The corrections applied to various 
data sets is shown in Table.~ 3.
Notice the correction for our observations at VBT is $<$ 0.01m
and falls within the uncertainty in the observations.

We attempt to fit the data on intensities in R-band to a time evolution 
defined by Eq.~1.
 
\begin{eqnarray}
I(t)&=&I(t_1) (t/t_1)^{\alpha_1} \quad \quad \hbox{for} \> \,t<t_1 \nonumber
 \\
I(t)&=&gI(t_1)(t/t_1)^{\alpha_1}+(1-g)I(t_1)(t/t_1)^{\alpha_2}\,\,\hbox{for}
 \,\> t_{1}\le t <t_{2}\nonumber \\
I(t)&=&\{g(t_2/t_1)^{\alpha_1}+(1-g)(t_2/t_1)^{\alpha_2}\} I(t_1)(t/t_2)
^{\alpha_2}\,\, \hbox{for}\> \, t\ge t_2  
\end{eqnarray}
This form is motivated by the feature observed in the R-band light curve
at $t-t_{0} \simeq 4.5 \, \rm days$. This  might have been
generated by a major burst followed after a short interval by a minor burst,
each being represented by a power-law form with a slope $\alpha_1$ which at later
times steepens to a slope $\alpha_2$. The function given in Eq.~(1) is designed to
test this hypothesis.  
Notice that this function has 6 parameters $I(t_1)$, g, $t_1$, $t_2$, $\alpha_1$ 
and $\alpha_{2}$, two more than the functional form adopted by 
Jensen et al. (2000)
The values of the parameters and their $1\sigma$ errors which provide the best fit 
to the data along with the normalized 
$\chi^{2}$ are computed using 'Levenberg-Marquardt method' (Press et al. 1992) 
and are listed in Table.~4. Notice that our fit for very early and very
late times
coincides quite closely with the fit obtained by Jensen et al. (2000) and has
a $\chi^2_{65} =1.80$. 
Even though Jensen et al. (2000) obtain with a single break an excellent fit to their
own data,
the single-break form fails to reproduce the compilation of world data which have
a slightly closer
coverage around the break point at 4--5 days after the burst; the $\chi^2$ 
they quote for
the fit to the world data is 3.687 for $\nu = 88$. 
After applying the corrections given in Table.~3 to
the data in Table~2. we fit the single-break form in Eq.~1 of Jensen et al. (2000)
and find that the $\chi^{2}_{64}$
reduces to 1.95 from its value of 2.85 for the same data points when no magnitude
corrections are applied. 
Thus with suitable corrections for normalizing different photometric data the
$\chi^{2}$ reduces to 1.8 ($\nu=65$) for two-break form and to 1.95 ($\nu =64$) for 
single-break form.
The improvement on the $\chi^{2}$ with double-break is significant at 38\% level as
seen from F-distribution test, although we can not rule out the fact that the high
$\chi^2$ may be due to a genuine variations due to inhomogeneities in the medium
surrounding the GRB host.

The motivation for the function given in Eq.~(1) is clearly
seen in Fig.~1.
The OT is fit with two events each of the type where a power-law decay sharply
steepens into another power-law several days after the burst.  
The light curve in the K-band reported by Rhoads \& Fruchter (2000) agrees broadly
with the fit we have obtained here for the first break (Eq~1.). If their curve
is simply extrapolated and compared with the data obtained by Jensen et al. (2000)
in the later and fainter epochs then one may find an apparent color evolution.
More accurate photometry with better
coverage is needed to confirm and characterize the possible existence of secondary
events which may be expected in some supernova-shock models 
(Meszaros, Rees and Weijers 1998). In case of GRB980228 
the re-brightening of the afterglow $\sim$3 weeks after the burst event and
the reddening of the spectrum have been hypothesized (Bloom et al. 1999) to be 
due to an underlying SN explosion  which triggered from the energy released
by a 'Collapsar' (MacFadyen \& Woosely 1999) that gave rise to the initial GRB event. 
Reichart(1999) also explained the color evolution in case of GRB980228 using
the supernova hypothesis.

To check if there are any such color changes during the evolution 
of GRB000301c we display in
Fig.~2. the behaviour of the light curve both in the B and in the R bands.
It is seen that the evolution of R and B magnitudes are nearly 
the same within the photometric uncertainties. This result has been reported
by Jensen et al. (2000) and Masetti et al. (2000) with a smaller sample, although
Rhoads \& Fruchter (2000) finds shift in R--K' color towards blue.
Running a F-test (Press et al. 1992)
gives $90\%$ probability that the two distributions are equal. The color B--R can be
fit with a function $ (B-R)=a(t-t_0)^{n}$ where, $a$ and $n$ are the two fitted 
parameters. Our best fit values are: $a =0.793\pm0.073$ and the slope 
$n=0.073 \pm0.074$ with a reduced $\chi^{2} =1.3$. This is consistent with
the achromatic behaviour of the B--R evolution.  
Hence chromatic behaviour is not established on the basis of B--R data and 
the suggestions of color variations are 
based on comparison of R-evolution with the K data by Rhoads \& Fruchter (2000).

As a closing remark to this brief paper we would like to add that
Vainu Bappu Observatory and the Uttar Pradesh State Observatory being located
at a longitude of $\sim 78^{\circ}$E in India could make the earliest set of
observations of the afterglow of GRB000301c and that excellent observational
opportunities have opened up at longitude $78^{\circ}57^{\prime}51^{\prime\prime}$E,
latitude $32^{\circ}46^{\prime}46^{\prime\prime}$N  with the
commissioning of a 2-m telescope at Hanle ($\sim$15000ft) in September 2000.

\acknowledgments{ Authors would like to thank the anonymous referee for
useful comments.
Mr J. S. Nathan is acknowledged for updating us on GCN
circulars during the observations. }

\newpage

\figcaption[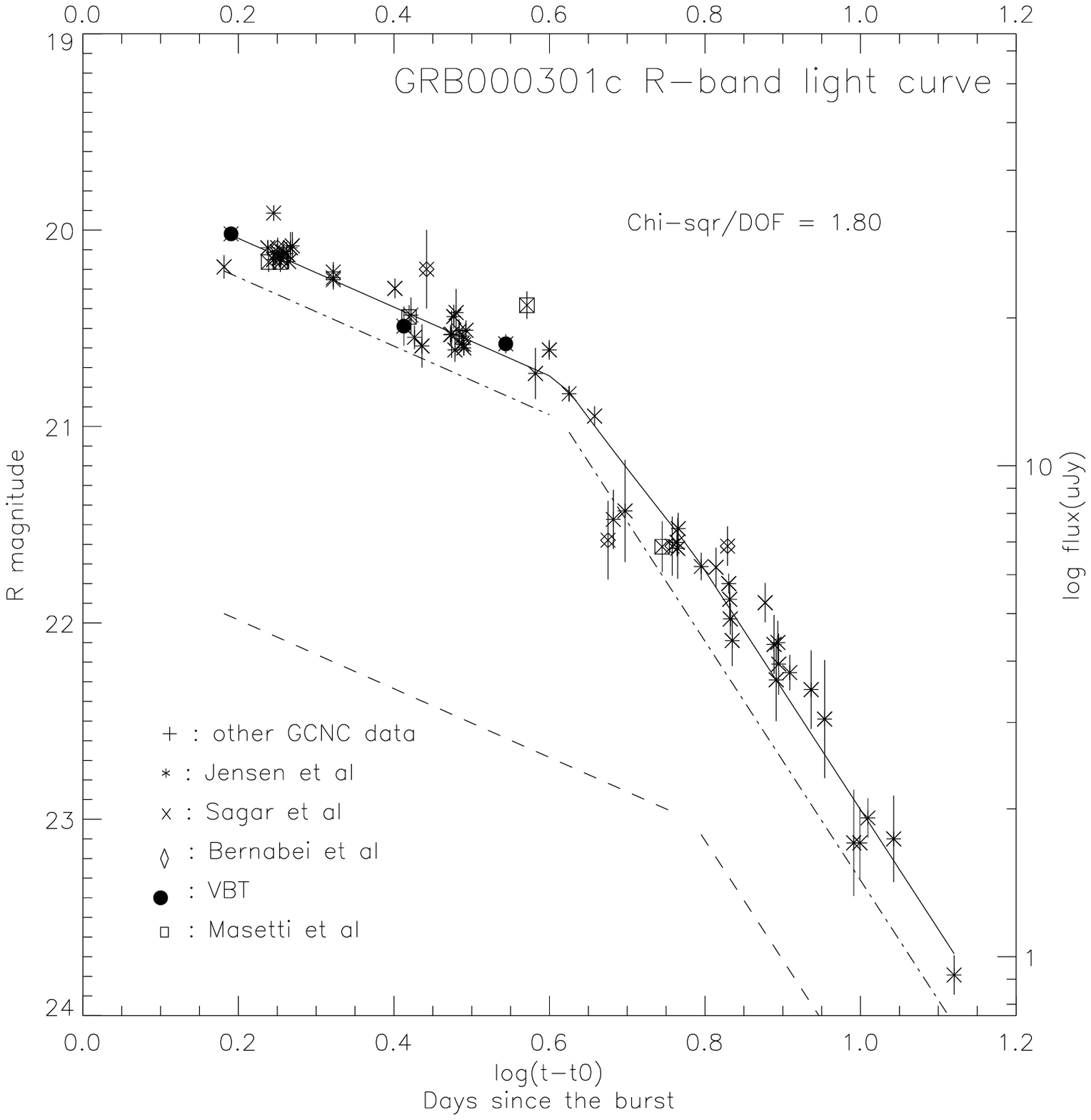]{GRB000301c R-band light curve: The dotted and
dashed lines represent the major and minor burst which add up to give
the light curve shown as a solid line. The data points from VBT are marked
by 'filled circles'.\label{fig1}}

\figcaption[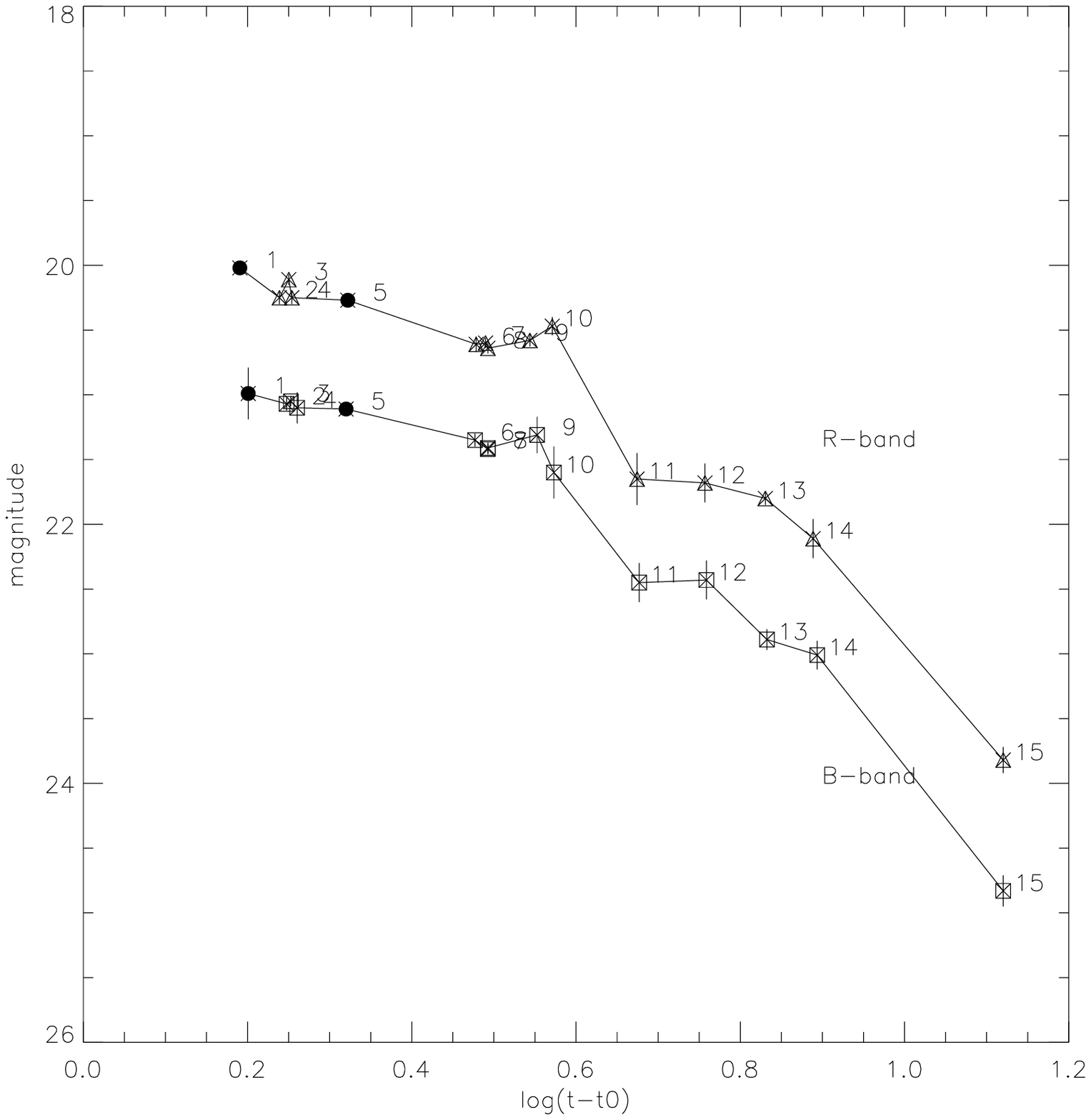]{GRB000301c light curve in B and R bands. 
The data points from VBT are marked by 'filled circles'. \label{fig2}}

\begin{deluxetable}{llllll}
\tablewidth{0pt}
\tablecaption{Log of the observations at VBT.\label{obs}}
\tablehead{\colhead{UT(days)} & \colhead{Filter} & \colhead{Total exp (sec)} 
& \colhead{Magnitude of OT} & \colhead{mag error} & \colhead{Remarks}}
\startdata
Mar 2.9618 & R      & 2400                 & 20.02     & 0.028 & 4 x 600sec summed \\
Mar 2.9986 & B      & 1200                 & 20.99     & 0.17  & single frame \\
Mar 3.9323 & R      & 1800                 & 20.45     & 0.12  & 3 x 600sec summed \\
Mar 3.9976\tablenotemark{b}

 & R      & 1200                 & 20.49     & 0.10  & 2 x 600sec summed \\
Mar 3.9694 & I      & 3000                 & 20.09     & 0.10  & 2 x 600s + 2 x 900s \\
Mar 4.9087 & R      & 3600                 & 20.57     & 0.05  & 6 x 600s summed \\
Mar 4.9799 & B      & 1800                 & 21.29     & 0.12  & single frame \\
Mar 4.9517 & I      & 1800                 & 19.96     & 0.05  & 3 x 600s summed \\
\enddata
\tablenotetext{b}{On Mar 3, between 2 set of R-band observations sky conditions varied and
therefore the frames are combined seperately. It may be noted that we have
considered the second set in our light curve plot.}
\end{deluxetable}

\begin{deluxetable}{llll}
\tablecolumns{4}
\tabletypesize{\small}
\tablecaption{Photometric data from literature \label{mag}}
\tablehead{ \colhead{UT(days)} & \colhead{R}
&\colhead{err(R)} & \colhead{Authors}}
\startdata

     2.930 &   20.42 &   0.06 &  Sagar et al.   \\
     2.962 &   20.02 &   0.028&  Bhargavi \& Cowsik     \\
     3.140 &   20.09 &   0.04 &  Jensen et al.   \\
     3.144 &   20.25 &   0.05 &  Masetti et al.      \\
     3.170 &   19.94 &   0.04 &  Fynbo et al.     \\
     3.170 &   20.15 &   0.04 &  Jensen et al.     \\
     3.190 &   20.11 &   0.04 &  Jensen et al.     \\
     3.191 &   20.11 &   0.05 &  Bernabei et al.     \\
     3.200 &   20.14 &   0.05 &  Jensen et al.     \\
     3.205 &   20.25 &   0.05 &  Masetti et al.     \\
     3.210 &   20.14 &   0.04 &  Jensen et al.     \\
     3.220 &   20.11 &   0.05 &  Jensen et al.     \\
     3.240 &   20.12 &   0.06 &  Jensen et al.     \\
     3.250 &   20.16 &   0.04 &  Jensen et al.     \\
     3.260 &   20.09 &   0.08 &  Jensen et al.     \\
     3.270 &   20.08 &   0.07 &  Jensen et al.     \\
     3.510 &   20.28 &   0.05 &  Garnavich et al.     \\
     3.510 &   20.27 &   0.04 &  Veilet \& Boer     \\
     3.510 &   20.24 &   0.05 &  Halpern et al.     \\
     3.930 &   20.53 &   0.05 &  Mohan et al.     \\
     3.998 &   20.49 &   0.1  &  Bhargavi \& Cowsik     \\
     4.038 &   20.53 &   0.06 &  Masetti et al.     \\
     4.052 &   20.46 &   0.09 &  Castro-Tirado et al.     \\
     4.079 &   20.57 &   0.06 &  Gal-Yam et al.     \\
     4.140 &   20.59 &   0.11 &  Jensen et al.     \\
     4.178 &   20.22 &   0.20 &  Bernabei et al.     \\
     4.380 &   20.56 &   0.05 &  Garnavich et al.     \\
     4.390 &   20.53 &   0.12 &  Jensen et al.     \\
     4.410 &   20.44 &   0.06 &  Jensen et al.     \\
     4.420 &   20.61 &   0.06 &  Jensen et al.     \\
     4.430 &   20.42 &   0.12 &  Jensen et al.     \\
     4.458 &   20.54 &   0.06 &  Mujica et al.     \\
     4.480 &   20.58 &   0.04 &  Jensen et al.     \\
     4.490 &   20.54 &   0.04 &  Jensen et al.     \\
     4.500 &   20.61 &   0.04 &  Halpern et al.     \\
     4.500 &   20.60 &   0.04 &  Jensen et al.     \\
     4.520 &   20.51 &   0.05 &  Jensen et al.     \\
     4.909 &   20.58 &   0.048&  Bhargavi \& Cowsik     \\
     5.135 &   20.47 &   0.07 &  Masetti et al.     \\
     5.230 &   20.73 &   0.13 &  Jensen et al.     \\
     5.390 &   20.61 &   0.05 &  Jensen et al.     \\
     5.630 &   20.86 &   0.04 &  Veilet \& Boer     \\
     5.960 &   21.18 &   0.05 &  Mohan et al.     \\
     6.145 &   21.60 &   0.20 &  Bernabei et al.     \\
     6.220 &   21.50 &   0.15 &  Fruchter et al.     \\
     6.390 &   21.43 &   0.26 &  Jensen et al.     \\
     6.968 &   21.70 &   0.13 &  Masetti et al.     \\
     7.135 &   21.63 &   0.15 &  Bernabei et al.     \\
     7.220 &   21.59 &   0.07 &  Jensen et al.     \\
     7.230 &   21.62 &   0.155&  Jensen et al.     \\
     7.240 &   21.52 &   0.08 &  Jensen et al.     \\
     7.650 &   21.74 &   0.07 &  Veilet \& Boer     \\
     7.930 &   21.95 &   0.1  &  Sagar et al.     \\
     8.157 &   21.63 &   0.1  &  Bernabei et al.     \\
     8.180 &   21.80 &   0.05 &  Jensen et al.     \\
     8.200 &   21.88 &   0.09 &  Jensen et al.     \\
     8.210 &   21.98 &   0.08 &  Jensen et al.     \\
     8.250 &   22.09 &   0.13 &  Jensen et al.     \\
     8.950 &   22.13 &   0.1  &  Sagar et al.     \\
     9.150 &   22.11 &   0.15 &  Jensen et al.     \\
     9.200 &   22.29 &   0.21 &  Jensen et al.     \\
     9.240 &   22.10 &   0.11 &  Jensen et al.     \\
     9.260 &   22.21 &   0.155&  Jensen et al.     \\
     9.520 &   22.28 &   0.09 &  Halpern et al.     \\
    10.050 &   22.34 &   0.20 & Jensen et al.     \\
    10.400 &   22.49 &   0.30 & Jensen et al.     \\
     11.21 &   23.12 &   0.27 & Jensen et al.     \\
     11.39 &   23.12 &   0.18 & Jensen et al.     \\
     11.63 &   23.02 &   0.1  & Veillet \& Boer     \\
     12.44 &   23.10 &   0.22 & Jensen et al.     \\
     14.60 &   23.82 &   0.1  & Veillet \& Boer     \\
     ...   & B--R \tablenotemark{a}   &   e(B--R) &   ...  \\
     2.975   &   0.9686  &   0.207 &  Bhargavi \& Cowsik \\
     3.162   &   0.82    &   0.07  &  Masetti et al. \\
     3.195   &   0.94    &   0.056 &  Jensen et al. \\
     3.218   &   0.85    &   0.13  &  Masetti et al. \\
     3.51    &   0.84    &   0.06  &  Veillet \& Boer \\
     4.415   &   0.74    &   0.078 &  Jensen et al. \\
     4.51    &   0.82    &   0.056 &  Jensen et al. \\
     4.52    &   0.77	 &   0.06  &  Halpern et al. \\
     4.9375  &   0.734	 &   0.14  &  Bhargavi \& Cowsik \\
     5.1435  &   1.13    &   0.21  &  Masetti et al. \\
     6.149   &   0.80    &   0.25  &  Masetti et al. \\
     7.137   &   0.75    &   0.212 &  Masetti et al. \\
     8.195   &   1.09    &   0.09  &  Jensen et al. \\
     9.195   &   0.9     &   0.186 &  Jensen et al. \\
     14.6    &   1.01	 &   0.12  &  Veillet \& Boer \\

\enddata
\tablenotetext{a}{B-R colors used in Fig.~2.}
\end{deluxetable}

\begin{deluxetable}{ll}
\tablecolumns{2}
\tablecaption{magnitude corrections.\label{mag_sh}}
\tablehead{ \colhead{data set} & \colhead{magnitude shift}} 
\startdata
vbt & -0.009 \\
Masetti & 0.088 \\
Bernabei & -0.02 \\
Sagar & 0.232 \\
Other GCN data & 0.03 \\
\enddata
\end{deluxetable}
 
\begin{deluxetable}{lll}
\tablecolumns{3}
\tablecaption{Fit parameters.\label{fit_val}}
\tablehead{\colhead{parameter} & \colhead{value} & \colhead{error}} 
\startdata
$I(t_1)$ & 14.88\,$\mu$Jy & 1.28\,$\mu$Jy \\
g  & 0.169 & 0.30\\
$t_1$ & 4.12\,days &0.30\,days \\
$t_2$ & 6.07\,days & 0.44\,days \\
$\alpha_{1}$ & -0.70 & 0.07 \\
$\alpha_{2}$ & -2.44 & 0.29 \\
$\chi^2_{65}$ & 1.80 &  --\\
goodness-of-fit & $2.65\times10^{-4}$ &-\\
\enddata
\end{deluxetable}


\begin{figure}
\figurenum{1}
\plotone{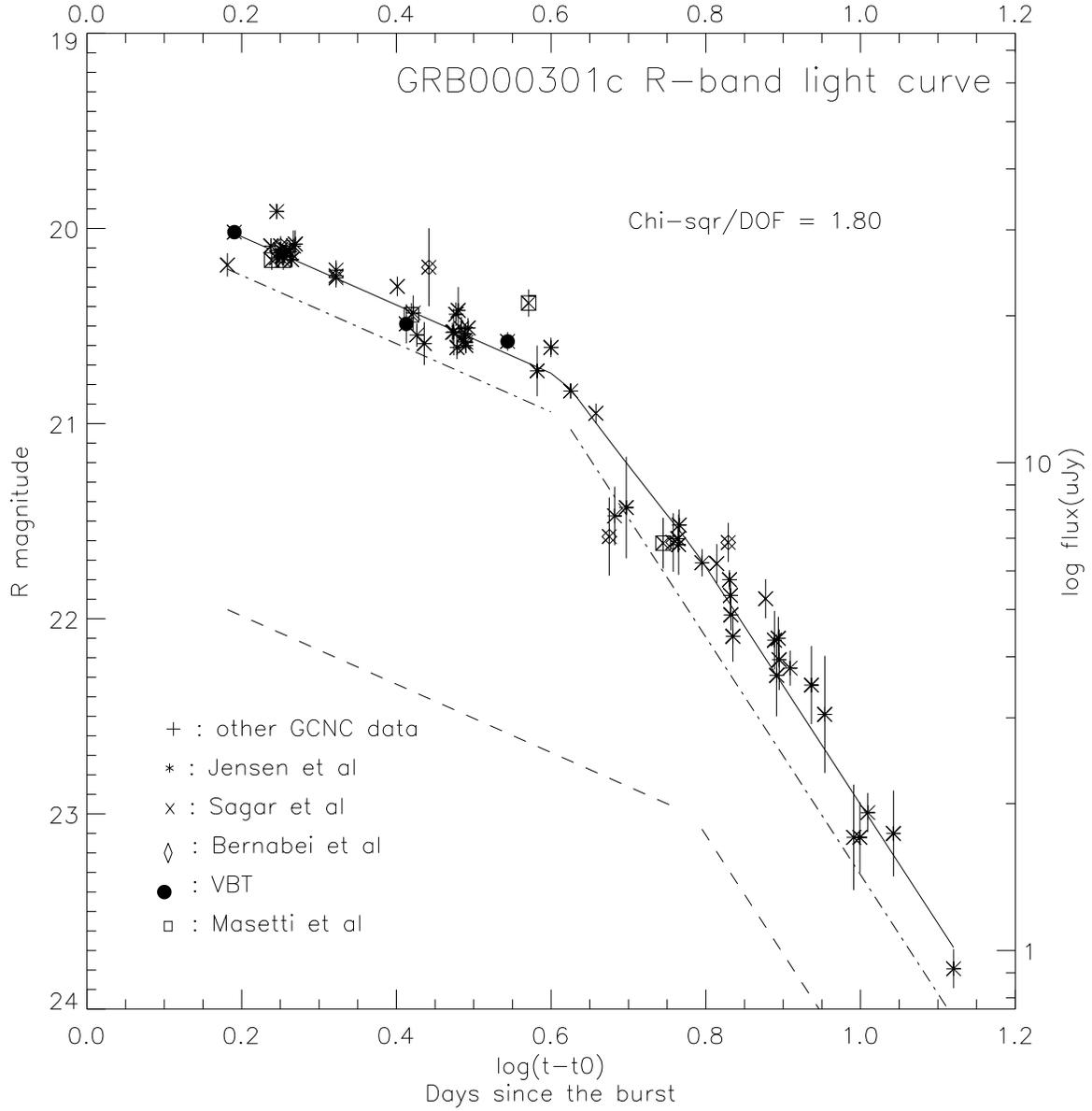}
\caption{GRB000301c R-band light curve: The dotted and
dashed lines represent the major and minor burst which add up to give
the light curve shown as a solid line. The data points from VBT are
marked by 'filled circles'. \label{fig1}}
\end{figure}

\begin{figure}
\figurenum{2}
\plotone{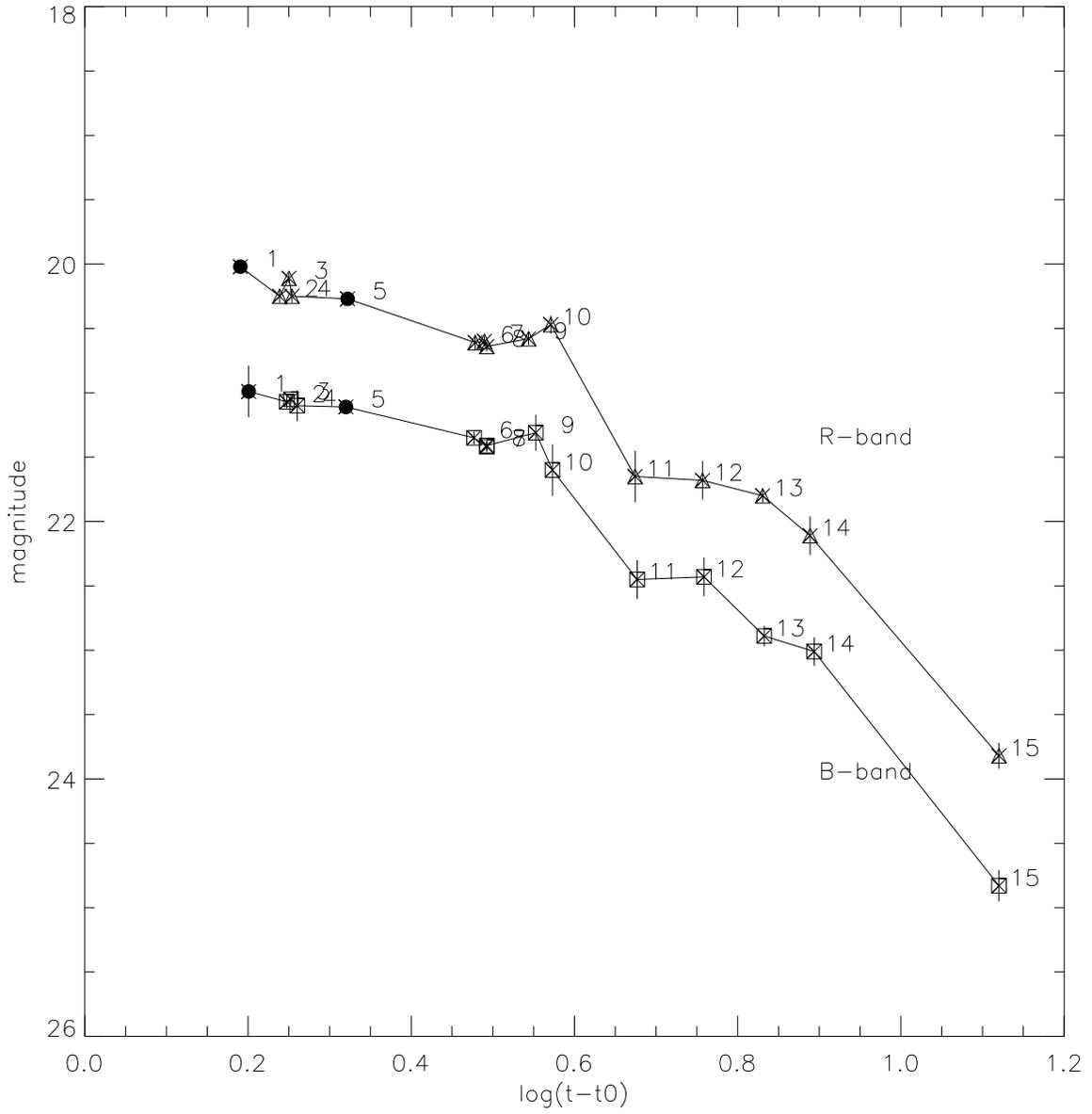}
\caption{ GRB000301c light curve in B and R bands. The data points from
VBT are marked by 'filled circles'.\label{fig2}}
\end{figure}
\end{document}